\documentclass[conference]{IEEEtran}
\IEEEoverridecommandlockouts
\usepackage{cite}
\usepackage{amsmath,amssymb,amsfonts}
\usepackage{algorithmic}
\usepackage{graphicx}
\usepackage{textcomp}
\usepackage{xcolor}
\usepackage{pifont}
\usepackage{url}
\usepackage{booktabs}
\usepackage{enumitem}
\usepackage{xspace}
\newcommand{\tool}{REEF\xspace}

\usepackage[many]{tcolorbox}
\usepackage{bbding}
\newcommand{\CBrush}{\textcolor[RGB]{84,130,53}{\Checkmark}}
\newcommand{\XBrush}{\textcolor[RGB]{176,35,24}{\XSolidBrush}}
\newcommand{\TriUp}{\textcolor[RGB]{0,112,192}{$\Diamond$}}

\newcommand{\parh}[1]{\noindent\textbf{#1}}
\newcommand{\F}{Fig.}

\newcommand{\T}{Table}
\renewcommand{\S}{Sec.}

\newcommand{\finding}[2]{
  \smallskip
  \smallskip
\begin{tcolorbox}[width=\linewidth,boxrule=0pt,top=1pt, bottom=1pt, left=1pt,right=1pt, colback=gray!20,colframe=gray!20]
\textbf{Finding #1:} 
{#2}
\end{tcolorbox}}

\def\BibTeX{{\rm B\kern-.05em{\sc i\kern-.025em b}\kern-.08em
    T\kern-.1667em\lower.7ex\hbox{E}\kern-.125emX}}
\begin{document}

\title{REEF: A Framework for Collecting Real-World Vulnerabilities and Fixes
}

\author{
    \IEEEauthorblockN{Chaozheng Wang\IEEEauthorrefmark{4}\IEEEauthorrefmark{5}, Zongjie Li\IEEEauthorrefmark{3}\IEEEauthorrefmark{5}\thanks{\IEEEauthorrefmark{5} These authors contribute equally.}, Yun Peng\IEEEauthorrefmark{2}, Shuzheng Gao\IEEEauthorrefmark{2}, Sirong Chen\IEEEauthorrefmark{4}}

    \IEEEauthorblockN{Shuai Wang\IEEEauthorrefmark{3}, Cuiyun Gao\IEEEauthorrefmark{4}\IEEEauthorrefmark{1}\thanks{* Cuiyun Gao is the corresponding author.}, Michael R. Lyu\IEEEauthorrefmark{2}}
    \IEEEauthorblockA{\IEEEauthorrefmark{4} School of Computer Science and Technology, Harbin Institute of Technology, Shenzhen, China}
    \IEEEauthorblockA{\IEEEauthorrefmark{3} Department of Computer Science and Engineering, Hong Kong University of Science and Technology, Hong Kong, China}
    \IEEEauthorblockA{\IEEEauthorrefmark{2} Computer Science and Engineering Department, The Chinese University of Hong Kong, Hong Kong, China}

    \IEEEauthorblockA{\{wangchaozheng, 190110916\}@stu.hit.edu.cn,\{zligo, shuaiw\}@cse.ust.hk}

    \IEEEauthorblockA{\{ypeng, szgao23, lyu\}@cse.cuhk.edu.hk, gaocuiyun@hit.edu.cn}
}

\maketitle

\begin{abstract}

Software plays a crucial role in our daily lives, and therefore the quality and security of software systems have become increasingly important. However, vulnerabilities in software still pose a significant threat, as they can have serious consequences.
Recent advances in automated program repair have sought to automatically detect and fix bugs using data-driven techniques. Sophisticated deep learning methods have been applied to this area and have achieved promising results. However, existing benchmarks for training and evaluating these techniques remain limited, as they tend to focus on a single programming language and have relatively small datasets.
Moreover, many benchmarks tend to be outdated and lack diversity, focusing on a specific codebase. Worse still, the quality of bug explanations in existing datasets is low, as they typically use imprecise and uninformative commit messages as explanations.

To address these issues, we propose an automated collecting framework \tool to collect \textbf{RE}al-world vuln\textbf{E}rabilities and \textbf{F}ixes from open-source repositories. We focus on vulnerabilities since they are exploitable and have serious consequences. We develop a multi-language crawler to collect vulnerabilities and their fixes, and design metrics to filter for high-quality vulnerability-fix pairs. Furthermore, we propose a neural language model-based approach to generate high-quality vulnerability explanations, which is key to producing informative fix messages.
Through extensive experiments, we demonstrate that our approach can collect high-quality vulnerability-fix pairs and generate strong explanations. The dataset we collect contains 4,466 CVEs with 30,987 patches (including 236 CWE) across 7 programming languages with detailed related information, which is superior to existing benchmarks in scale, coverage, and quality. Evaluations by human experts further confirm that our framework produces high-quality vulnerability explanations.

\end{abstract}


\section{Introduction}
\label{sec:intro}

Software powers an increasing number of critical systems in the modern world, from people's daily life applications~\cite{jadhav2022evolution} to critical industrial scenarios~\cite{li2017fourth} or even military systems~\cite{wang2020investigation}. However, software often contains 
bugs that can cause substantial losses~\cite{4267025}. In this paper, we focus on vulnerabilities since they are a more severe type of bug related to software security that may have serious consequences. The Common Vulnerabilities and Exposures (CVE) database tracks publicly disclosed cybersecurity vulnerabilities and the number of reported CVEs has been growing rapidly over the past decade~\cite{mitre}, highlighting the scale and significance of this problem; this growth is largely attributed to the increasing number of open-source software applications, as well as the increasing sophistication of cyber-attacks and vulnerability discovery methods.
To combat the potential risk in real-world software, various approaches have been proposed to detect and fix bugs. 

Traditionally, static analysis tools have been used to analyze source code and detect potential bugs. A large number of mature tools such as Coverity~\cite{coverity} have been maintained and improved for several decades, applying to real-world projects to discover potential flaws. More recently, query-based code search tools like CodeQL~\cite{codeql-intro} and Semgrep~\cite{semgrep} have enabled developers to detect bugs by writing semantic queries to explore code bases. The core idea behind the tools is to provide basic analysis capabilities to developers and let them write queries to detect bugs. The effectiveness of these tools depends on the quality of the queries~\cite{10.1145/3548606.3563527}, which keeps the detection process both flexible and scalable.

Besides the traditional static analysis and query-based code search tools, data-driven approaches~\cite{DBLP:journals/tse/ChenKTPPM21,DBLP:conf/sigsoft/ZhuSXZY0Z21,DBLP:conf/sigsoft/XiaZ22,DBLP:conf/icse/Li0N20,DBLP:conf/issta/LutellierPPLW020,DBLP:conf/icse/JiangL021,DBLP:conf/icse/YeMM22,DBLP:conf/icse/Li0N22,DBLP:journals/corr/abs-2302-01857,DBLP:journals/corr/abs-2306-01394} are also promising for detecting and fixing bugs.
Different from analyzing the source code with predefined rules along with complex analysis techniques (e.g., dataflow analysis and control flow analysis),
data-driven approaches leverage large-scale data from open-source communities such as GitHub to learn the patterns of bugs and fixes. Techniques such as zero-day detector~\cite{cui2007shieldgen} generate candidate patches and validate them to find viable fixes. Other data-driven methods learn developer-written patches from open-source repositories and apply them to repair new bugs.  

The advantages of the data-driven approaches are two-fold. First, they are easier to implement than traditional static analysis tools. The traditional tools require complex analysis techniques to analyze the source code, which is time-consuming and hard to scale. For example, when facing a wide variety of software vulnerabilities, professional developers must carefully design appropriate detection rules or techniques~\cite{coverity,codeql-intro}. 
In contrast, the data-driven approaches only require large-scale data from open-source repositories, which is easy to obtain and scale. 
Second, they are more flexible than the traditional tools. The traditional tools are usually language-agnostic and designed for specific types of bugs, which are hard to extend to other programming languages or other types of bugs~\cite{juliet,defects4j,mu2018understanding}.
The data-driven approaches are more flexible, as they can be easily extended to other types of bugs by learning the patterns of bugs and fixes from the large-scale data.

Data-driven approaches for automated bug detection and repair rely on the availability of large-scale, high-quality datasets. The efficacy of these techniques is directly dependent on the precision and comprehensiveness of the data used to train machine learning models. Specifically, accurately identifying bug locations and patches can bolster the precision of the trained models, improving the accuracy of bug detection and fixing. Furthermore, incorporating metadata on bug types and commit messages provides models with more granular information about bugs, which can enhance their performance on detection and repair tasks~\cite{DBLP:conf/emnlp/BaiZB0W0021}. This additional context may also aid developers during debugging and code review by giving them more targeted insights. 

However, current datasets for data-driven software analysis have several key limitations that hamper their effectiveness. First, the granularity of most datasets is at the function level, 
lacking precise records of bug locations and fixes. In reality, bugs often span multiple levels of abstraction and a single CVE may impact various, disparate parts of a codebase~\cite{nappa2015attack}. Thus, there is a shortage of systematically curated, real-world vulnerability data. 
Second, metadata about bug types is often inaccurate or imprecise. Currently, bug types are typically inferred from commit messages, which can be inaccurate and even sometimes wrong~\cite{liu2018nmtbmessage}. Such an error-prone process would possibly yield incorrect or misleading labels.   
Third, most current datasets~\cite{zhou2019devign} are outdated, failing to capture the latest state of constantly evolving software systems. Bugs that were previously patched may be reintroduced, tending to render existing records obsolete.

\parh{Technical Challenges and Solutions.}~We aim to develop a framework for automatically collecting and curating high-quality code snippets containing vulnerabilities, fixes, locations, their types, and messages from open-source repositories. It is thus required to prepare a large dataset incorporating this information to gain insights into real-world bugs and fixes, facilitating further research and applications.
To achieve the above goals, our approach comprises three steps:
\ding{202}  Newest CVE capture: current datasets focus on simple and outdated vulnerability patterns without clear location information as well as the patch. To develop a comprehensive dataset that is close to real-world situations, we primarily focus on the recently revealed CVEs that share clear fix log and location information. This results in a dataset with about eighteen thousand CVEs that are recently revealed with documented fixes.
\ding{203} Large Language Model (LLM)-based explanation with human agreement: Current datasets adopt the commit information as the comment for the vulnerability content, which is proved to be unreliable~\cite{10.1145/1882291.1882308,10.1145/3510003.3510205}. 
Considering the powerful capability of LLM in code understanding \cite{peng2023domain, gao2023constructing}, we leverage the large language model to automatically comprehend the bug patterns and fixes from the CVEs and use them as the additional message. To ensure the quality of the mined results, we carefully design the prompt system with a pilot study. Moreover, we conduct a human agreement study to evaluate the mined results.
\ding{204} Analysis of the collected dataset:
we conduct further analysis of the dataset to understand the characteristics of real-world bugs and fixes, providing details on each data point to benefit future studies or applications.  

Our contributions are summarized as follows:
\begin{itemize}
\item We propose \tool, a framework to mine up-to-date, real-world vulnerabilities automatically. Incorporated with the corresponding fix patches from CVEs, we have collected a large-scale dataset with {30,987} bug location, type, and fix information. {Our dataset consists of a wide variety of vulnerabilities across various languages, platforms, and granularity.}
\item We employ large language models to generate explanatory messages for the CVEs, supplementing unreliable commit information. We carefully design prompts and conduct a human evaluation to ensure message quality.
\item We conduct an extensive analysis of the collected dataset and provide detailed insights into real-world vulnerabilities and fix characteristics to guide future research and applications.
\item We have publicly released the code for our REEF tool on GitHub at \textit{\url{https://github.com/ASE-REEF/REEF-script}}, along with the vulnerability data we have collected, which is also available on GitHub at \textit{\url{https://github.com/ASE-REEF/REEF-data}}.
\end{itemize}

\section{Related work}
\label{sec:related}
In this section, we introduce the related work from three threads including automated program repair, static analysis tools, and large language models for code, respectively.
\subsection{Automated Program Repair}
\label{sec:apr}
Automatic program repair (APR) has garnered significant attention in recent years as a crucial approach to enhancing software reliability. Various techniques have been developed within the APR domain, including template-based~\cite{DBLP:conf/issta/LiuK0B19,DBLP:journals/ese/KoyuncuLBKKMT20}, search-based~\cite{DBLP:conf/issta/JiangXZGC18,DBLP:journals/tse/GouesNFW12}, constraint-based~\cite{DBLP:conf/kbse/PeiWFNM11,DBLP:journals/tse/0001FNWMZ14}, and learning-based approaches~\cite{DBLP:journals/tse/ChenKTPPM21,DBLP:conf/sigsoft/ZhuSXZY0Z21,DBLP:conf/sigsoft/XiaZ22,DBLP:conf/icse/Li0N20,DBLP:conf/issta/LutellierPPLW020,DBLP:conf/icse/JiangL021,DBLP:conf/icse/YeMM22,DBLP:conf/icse/Li0N22,DBLP:journals/corr/abs-2302-01857,DBLP:journals/corr/abs-2306-01394}. Among these categories, learning-based methods have achieved the greatest success and become the most popular in recent years. 
SequenceR~\cite{DBLP:journals/tse/ChenKTPPM21} combines LSTM encoder-decoder architecture with copy mechanism for program repair. DLFix~\cite{DBLP:conf/icse/Li0N20}
uses a tree-based RNN to capture the structure of the source code and learn code transformations. CoCoNuT~\cite{DBLP:conf/issta/LutellierPPLW020} uses ensemble learning on the combination of different networks to automatically fix bugs in multiple programming languages, separating the context and buggy lines in NMT-based APR. CURE~\cite{DBLP:conf/icse/JiangL021} integrates pre-trained programming language models and significantly improves repair quality. Recoder~\cite{DBLP:conf/sigsoft/ZhuSXZY0Z21} uses a syntax-guided edit decoder to guide the generation of syntactically correct repair patches. RewardRepair~\cite{DBLP:conf/icse/YeMM22} employs execution-based backpropagation to enhance the compilation rate of patches generated by NMT-based APR approaches. DEAR~\cite{DBLP:conf/icse/Li0N22} generate multi-hunk, multi-statement fix patches with a divide-and-conquer strategy. AlphaRepair~\cite{DBLP:conf/sigsoft/XiaZ22} utilizes a large pre-trained code model and generates patches in a fill-in-the-blank way. Zhong et al.~\cite{DBLP:conf/kbse/ZhongGALLG022}  build a standard benchmark dataset and an extensive framework tool to mitigate threats for comparison in program repair. Xia et al.~\cite{DBLP:journals/corr/abs-2210-14179} evaluate the effectiveness of LLMs on program repair. KNOD~\cite{DBLP:journals/corr/abs-2302-01857} incorporates domain knowledge to guide patch generation in a direct and comprehensive way. TypeFix~\cite{DBLP:journals/corr/abs-2306-01394} is a prompt-based approach with fix templates incorporated for repairing Python type errors.

\subsection{Static Analysis Tools}
\label{sec:sast}

Static analysis tools analyze source code without executing the program to detect potential bugs and vulnerabilities. They codify definitions of unsafe coding patterns and scan codebases to identify matches. Over the past decade, static analysis has become a popular approach for detecting vulnerabilities in software~\cite{chess2004static}, and techniques including data flow analysis~\cite{bodden2012inter,liu2023exploring}, typestate analysis~\cite{fink2008effective}, type inference~\cite{DBLP:conf/icse/PengGLG0ZL22} and specific pointer analysis~\cite{smaragdakis2015pointer} have been developed to improve the precision and recall of bug detection.

Recently, query-based static analysis tools like CodeQL~\cite{codeql-intro} have gained increasing attention from both academia and industry. These tools codify vulnerability patterns as SQL-like queries, facilitating knowledge sharing and reuse across entities and software systems. Software is treated as data~\cite{yamaguchi2014modeling}, with programs parsed into hierarchical representations, often stored in databases.
Unlike traditional static analysis tools~\cite{coverity,facebookinfer}, query-based tools primarily focus on parsing software into rich, query-friendly representations and rely on crowdsourced communities to continually develop queries addressing newly discovered vulnerabilities. Established query-based tools cultivate active communities and offer bounty programs~\cite{codeqlbounty} to encourage query contribution and improvement. In turn, these communities help enrich and refine queries to target vulnerabilities proliferating in real-world software. 

There is a huge effort to establish comprehensive benchmarks to evaluate the quality of analysis tools, which further helps find real-world vulnerabilities. However, existing benchmarks for evaluating static analysis tools typically use synthetic datasets. For instance, the Juliet~\cite{juliet} benchmark for C/C++ and the Defects4J~\cite{defects4j} in Java language. These follow prescribed patterns and quickly become outdated, unable to represent the complexity of real-world CVEs. Although built around the Common Weakness Enumeration (CWE) to provide reasonable, well-defined examples, they cannot capture the nuances of most vulnerabilities.  
Some works collect datasets from real CVEs and make complex processes to filter the suitable ones. For instance, Ruohonen~\cite{ruohonen2018empirical} targets Python and collects samples from popular repositories and Linares et al.~\cite{linares2017empirical} analyze Android apps. These narrow scopes limit the types of vulnerabilities and coding patterns addressed, impeding holistic analysis. However, these works usually focus on particular programming languages or codebases, lacking diversity.

While static analysis shows promise for detecting vulnerabilities at scale, evaluating tools remains challenging due to the lack of comprehensive benchmarks reflecting the diversity of real-world bugs. Real CVEs offer a rich source for dataset generation but are difficult and time-consuming to gather and curate. Progress in static analysis thus depends on developing datasets that mirror the heterogeneity of vulnerabilities in real code. Automated or semi-automated methods for collecting and labeling examples from a wide range of open-source repositories directly show potential for advancing research and practice in this crucial area of software security, which is the focus of our work.

\subsection{Large Language Models for Code}
\label{sec:llm-code}

Recently, significant advancements in SE research have been brought by Large Language Models (LLMs), which brought impressive improvements in a wide range of code-related tasks. One notable model is Incoder~\cite{DBLP:journals/corr/abs-2204-05999}, which employs a causal masking training objective to excel in code infilling and synthesis. Another popular model is Codex~\cite{DBLP:journals/corr/abs-2107-03374}, a sizable pre-trained code model introduced by OpenAI, which supports the Copilot service on various code-related tasks~\cite{li2022cctest}. The models recently released by OpenAI, such as ChatGPT~\cite{chatgpt} and GPT-4~\cite{GPT4}, are also pre-trained with source code data and show remarkable programming capabilities.  AlphaCode~\cite{DBLP:journals/corr/abs-2203-07814} has been specifically trained for generating code for programming competitions like Codeforces.
CodeCMR~\cite{yu2020codecmr} and IRGEN~\cite{li2022unleashing} are pre-trained models designed for low-level code on various code-related tasks. 
CodeGen~\cite{nijkamp2022codegen} is a large pre-trained model for multi-turn program synthesis with more than 16B parameters, while CodeGeeX~\cite{CodeGeeX} is a recently proposed open-source multilingual code generation model with 13 billion parameters. BigCode Project has developed and open sourced StarCoder~\cite{DBLP:journals/corr/abs-2305-06161} which contains 15.5B parameter. A recent work WizardCoder~\cite{DBLP:journals/corr/abs-2306-08568} is fine-tuned with Evol-Instruct and achieves state-of-the-art performance surpassing all existing open-source Code LLMs.

In-context learning (ICL)~\cite{dong2022survey,DBLP:conf/nips/BrownMRSKDNSSAA20} is a recent paradigm that enables LLMs to learn from just a few examples without fine-tuning. It concatenates a few input-output examples with the query question to form an input for the language model and get the prediction. Recently, there has been increasing interest in applying in-context learning to code-related tasks~\cite{DBLP:journals/corr/abs-2210-14179,DBLP:conf/icse-apr/PrennerBR22,nashidretrieval,DBLP:journals/corr/abs-2303-03012}. CEDAR~\cite{nashidretrieval} retrieves similar examples and constructs the demonstrations for assert generation and program repair. Synchromesh~\cite{DBLP:conf/iclr/PoesiaP00SMG22} retrieves few-shot examples by Target Similarity Tuning and samples programs using Constrained Semantic Decoding. A recent work~\cite{DBLP:journals/corr/abs-2304-07575} empirically studies the impact of three demonstration construction factors on in-context learning in code intelligence tasks. Geng et al.~\cite{geng2024large} enhance in-context learning for multi-intent code comment generation by selecting similar examples and re-ranking the output candidates. Ahmed et al.~\cite{DBLP:journals/corr/abs-2304-06815} propose to incorporate static analysis results into the in-context prompt to code summarization.

\section{Workflow}
\label{sec:workflow}

\begin{figure*}[!htbp]
    \includegraphics[width=1.0\linewidth]{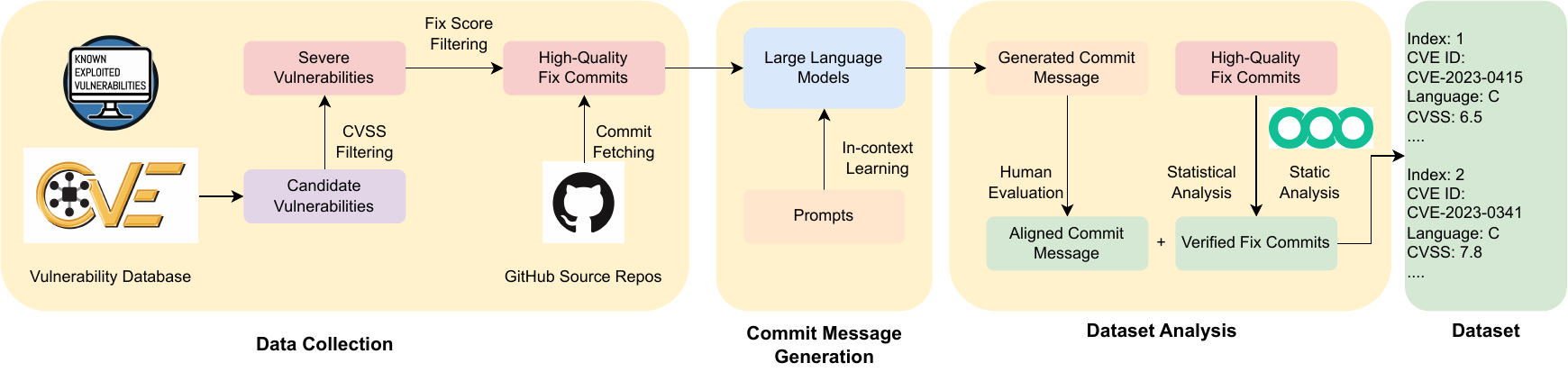}
    \caption{The pipeline of our \tool for gathering vulnerabilities, enriching data, and analyzing the dataset.}
    \label{fig:workflow}
\end{figure*}

\F~\ref{fig:workflow} presents our pipeline for collecting real-world vulnerabilities and constructing the dataset. Our pipeline consists of three steps:    
\ding{202} CVE capturing and collection. To address the limitations of current benchmarks, we gather recent real-world vulnerabilities from CVEs that have been newly disclosed, including related bug reports and source code. We use an automated crawler to collect the latest CVEs and then filter out irrelevant ones based on metrics like CVSS scores.  
\ding{203} LLM message supplementation. To account for uneven commit message quality and potential bias, we leverage the ability of large language models~\cite{pearce2021examining} to generate vulnerability explanations for each commit. Following standard bug description guidelines~\cite{googlebugformat}, we design a prompt for the vulnerability explanations using advanced LLM, enabling the model to formulate consistent, unified messages through contextual learning. 
\ding{204} Dataset analysis. We introduce metrics to assess our dataset's quality and compare it with current benchmarks to determine effectiveness in supporting existing tools. We analyze the generated messages and compare them with committed information to evaluate our approach. Finally, we conduct human studies to assess the quality of the generated messages.

To evaluate the effectiveness of our approach to collect code vulnerability and the quality of our proposed dataset, we investigate the following three research questions (RQs):
\begin{enumerate}[label=\bfseries RQ\arabic*:,leftmargin=.5in]
    \item What is the advantage of our dataset compared to existing benchmarks?
    \item To what extent the prompt design affects the generated message?
    \item How are the generated messages in alignment with experts?
\end{enumerate}

Specifically, we analyze our collected dataset and compare it with current benchmarks to explore whether it is effective and diverse in RQ1 (in Section \ref{subsec:rq1}). 
As we use the code understanding ability of LLM, in RQ2 we further study how the prompt would affect the performance of generated bug explanation compared to the commit information in Section \ref{subsec:rq2}. 
Finally in RQ3, by comparing the patches we collected as well as the generated message, we study to what extent humans are in agreement with the generated code explanation as discussed in Section \ref{subsec:rq3}.

Notably, we focus on a wide variety of vulnerabilities across various languages and platforms, not only including the 
commonly-seen pattern bugs occur at the function level or statement level, but also the complex vulnerabilities across multiple files and functions,
which we believe is more challenging for the current tools to detect and can better benefit the community.

\subsection{Data Collection}
\label{subsec:data-collection}

Creating an exhaustive dataset including real-world code snippets with vulnerabilities, fixes, locations, and types is challenging, let alone including vulnerability explanations. To develop a dataset reflecting real-world scenarios, we first gather real-world vulnerabilities from multiple sources, including the NVD database and CVE list maintained by Mend~\cite{whitesource}, which is a comprehensive open-source vulnerabilities database from hundreds of both popular and under-the-radar community resources. Users can also specify additional sources as needed. 
If a vulnerability has a clear report, proof-of-concept, and publicly available source code before and after fixing, we collect the related bug reports and source code, and store them in our raw dataset. Notably, it would be possible that a single vulnerability may be linked to multiple commits and files; we gather all related commits and files accordingly.

As shown in~\F~\ref{fig:workflow}, we design a filter to remove less severe vulnerabilities from the raw dataset. We first eliminate those with a low CVSS score which indicates relatively low
impact and damage potential. We then assign a ``fix score'' based on the number of related commits and a weighted score to each commit based on the number of files modified. Vulnerabilities with
low ``fix scores'' are excluded from the final dataset.

To expand the dataset's potential applications and adapt it for various downstream tasks, we incorporate disclosure date information for each item. Specifically, we keep the complete CVE name indicating when each vulnerability was disclosed and assigned a unique number. Users preferring more recent data can easily filter out vulnerabilities disclosed during a given time period.

\subsection{LLM Message Supplementation}
\label{subsec:llm-message}

To construct a comprehensive, informative dataset, vulnerability descriptions are crucial since they provide rationale and fix details, helping downstream tools better understand vulnerabilities. This is especially useful for data-driven methods, as natural language is easier to comprehend than code \cite{wang2022no}.
However, commit messages are not always available and may be uninformative,  biased~\cite{bird2009fair}, or misleading~\cite{liu2018nmtbmessage} in explaining vulnerabilities. Worse yet, the existing commit messages could be missing, impairing downstream approaches.
To address this, we leverage large language models to generate vulnerability explanations for each commit. Following standard bug description guidelines~\cite{googlebugformat}, we empirically design a prompt for the vulnerability explanations using advanced LLM, enabling the model to formulate consistent, unified messages through in-context learning.

After collecting the responses from the large language model, we further conduct a human inspection to evaluate whether 
the generated message is in alignment with experts and whether it is informative enough to explain the vulnerability. Only the generated message that is in agreement with the experts would be included in the final dataset.

We provide a complete list of field names for our collected dataset in \T~\ref{tab:item_desc}. The fields can be categorized into four groups: (1) metadata, (2) vulnerability information, (3) LLM-enhanced information, and (4) project information. The metadata describes the programming language and index for each data item. The vulnerability information contains CVE details documenting the real-world impact of the bugs. The LLM-enhanced messages are generated descriptions of the vulnerabilities.

\begin{figure}[t]
    \includegraphics[width=1.0\linewidth]{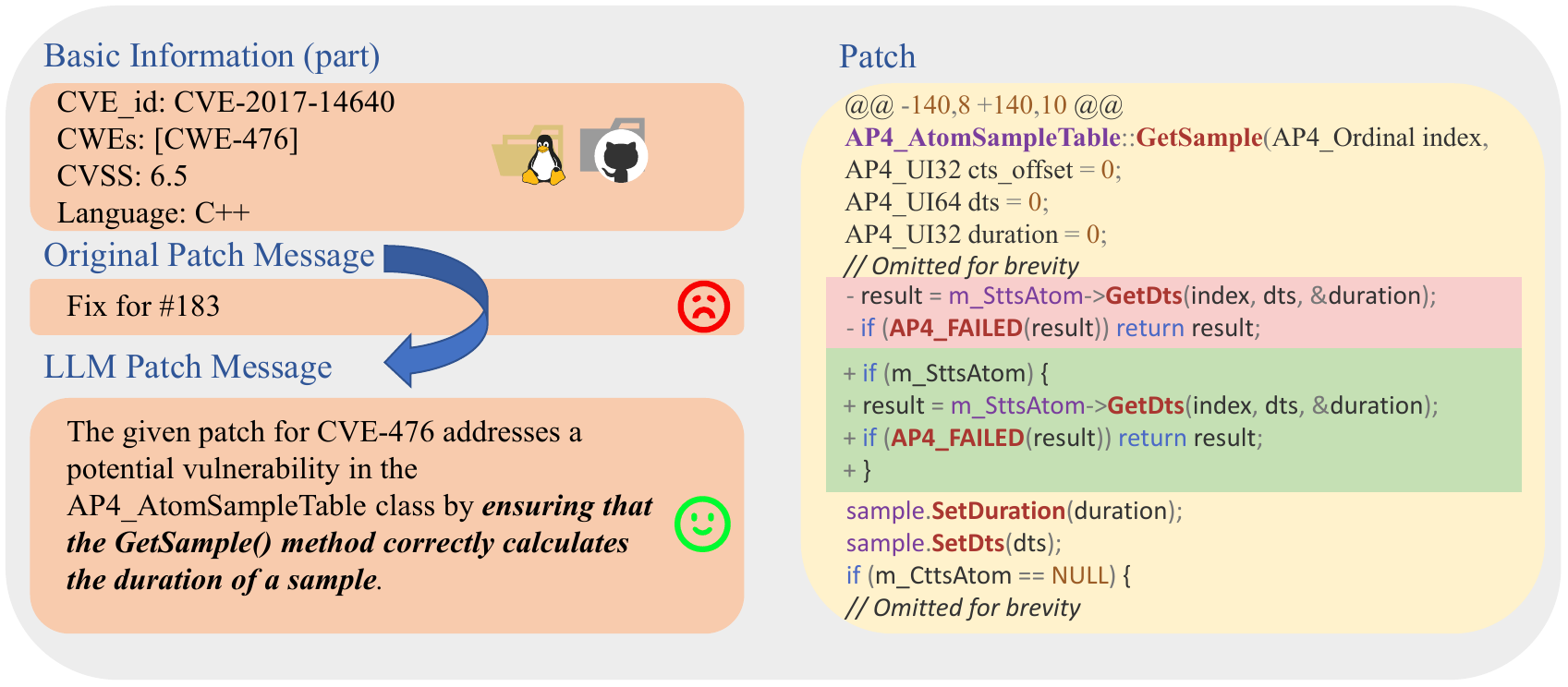}
    \caption{Example of data instance with enhanced patch messages.}
    \label{fig:dataitem}
\end{figure}

The project information stores data on the code repositories mined for bugs, including website details.
In summary, our dataset incorporates comprehensive metadata, security details, generated explanatory messages, and provenance information to enable in-depth analysis of real-world vulnerabilities. The multi-faceted data provides insights into vulnerability properties, remediation, language model performance, and codebase characteristics.

\begin{table}[t]
    \centering
    \caption{Description of Fields in the JSON File of Our Dataset.}
    \scalebox{0.9}{
    \begin{tabular}{cl}
    \toprule
        \textbf{Field Name in JSON} & \textbf{Description} \\
    \midrule
        \multicolumn{2}{c}{\textbf{Metadata}} \\
    \midrule
        \texttt{index} & The index of the instance in the dataset \\
    \midrule
        \texttt{language} & The programming language of the instance\\
    \midrule
        \multicolumn{2}{c}{\textbf{Vulnerability Information}} \\
    \midrule
        \texttt{cve\_id} & The identifier of CVE that the instance belongs to\\
    \midrule
        \texttt{cvss} & The CVSS score reflecting the severity of the CVE\\
    \midrule
        \texttt{cwes} & The CWEs, i.e., the common weaknesses in the CVE\\
    \midrule
        \multicolumn{2}{c}{LLM-enchanced Information} \\
    \midrule
        \texttt{llm\_message} & The commit message generated by LLMs\\
    \midrule
        \multicolumn{2}{c}{\textbf{Project Information}} \\
    \midrule
        \texttt{origin\_message} & The original commit message \\
    \midrule
        \texttt{url} & The URL of the commit on GitHub REST API \\
    \midrule
        \texttt{html\_url} & The URL that locates the webpage of the commit\\
    \midrule
        \texttt{raw\_url} & The URL of the source file changed in the commit\\
    \midrule
        \texttt{raw\_code} & The source code of the file changed in the commit\\
    \bottomrule
    \end{tabular}}
    \label{tab:item_desc}
\end{table}

\subsection{Dataset Analysis}
\label{subsec:dataset-analysis}

To rigorously evaluate our dataset's quality, we conduct an exhaustive analysis.  
Conceptually, we compare our dataset to current benchmarks across several axes: supported programming languages, {fixing} information integrity, granularity, data source, corpus size, and CWE coverage, {respectively}. A comparative analysis along these dimensions provides a holistic sense of relative strengths.
A systematic assessment of vulnerability type diversity examines coverage of the Common Weakness Enumeration (CWE). We posit that a dataset exhibiting a wider range of CWE types affords more comprehensive vulnerability modeling, with greater potential to generalize across systems. By comparing the number of CWE types, we gain quantitative insight into our dataset's diversity.
We further evaluate the challenge to current tools using Semgrep~\cite{semgrep}, a ubiquitous query-based static analysis tool, to detect vulnerabilities in our dataset. A higher proportion of vulnerabilities evading detection suggests greater resilience against existing methods, indicating the dataset poses a more formidable evaluation benchmark. Missed vulnerabilities point to remaining gaps in vulnerability modeling and detection that the dataset could help address through continued research.
Finally, we conduct an in-depth statistical analysis of our dataset itself to glean qualitative details that inform future work. Summary statistics on parameters like vulnerabilities' average severity and exploitability, bug fix length, and explanation shine a light on real-world characteristics underrepresented in synthetic data. A granular exploration of attributes exposes new problem dimensions beyond the capabilities of simplified synthetic benchmarks.

These complementary analyses, systematically connected through a logical pipeline, provide empirical evidence and qualitative characterization to demonstrate our dataset's diversity, challenge, and fidelity in emulating real-world scenarios. The rigor and depth of our evaluative approach underline the dataset's potential to serve as a foundation for future research advancing the state of the art in software vulnerability detection and automatic program repair.

\section{Evaluation}
\label{sec:evaluation}

\subsection{RQ1: What is the advantage of our dataset compared to existing benchmarks?}
\label{subsec:rq1}

\parh{Conceptual comparison.}~We first conduct a conceptual comparison between our selected dataset and current datasets, in which we compare the 
differences during the dataset collection process, including the multi-language support, fix information, location, related message, granularity, source, and size.

\begin{table*}[!thp]
    \caption{Comparasion our collected dataset and current benchmarks \CBrush, \TriUp,
          \XBrush\ denote support, partially support, and not support, respectively.}
          \centering
    \resizebox{0.85\linewidth}{!}{
    \begin{tabular}{lllllllll}
    \toprule
                & \# Sup language  & Fix information  & location      & Related Msg & Granularity & Source & Size & \# CWE Types \\ \hline
    Juliet-C++~\cite{juliet}  & 1         & \XBrush    &  \CBrush  & \XBrush   & statement-level  & Synthetic & 64,099 & 118\\ \hline
    Defects4J~\cite{defects4j}   & 1        & \CBrush     & \XBrush   & \XBrush   &  function-level & Synthetic & 835 & -\\ \hline
    LinuxFlaw~\cite{mu2018understanding} & 1   & \CBrush     & \XBrush   & \CBrush  &  multi-level &  CVE & 368  & 18\\ \hline
    FUNDED~\cite{9293321} & 4   & \CBrush     & \XBrush   & \XBrush  &  function-level & Synthetic \& CVE & 6561  & 54 \\ \hline
    Ours        & 7        & \CBrush     & \CBrush  & \CBrush     & multi-level    &  CVE & 30,987 & 236\\ 
    \bottomrule

    \end{tabular}
    }
    \label{tab:dataset}
    \end{table*}

As shown in Table~\ref{tab:dataset}, our dataset has the following advantages compared to current benchmarks:
(1) Detailed fix information. Compared to other benchmarks that only include fix patterns, our dataset contains detailed fix information, including the bug location and fix information. 
(2) Multi-language support. Different from the existing datasets that focus on specific languages, ours incorporate vulnerabilities in multiple languages, including C/C++, Java, C\#, and Python. 
(3) Multi-level granularity. Our dataset contains vulnerabilities at multiple levels, including function-level, statement-level, and expression-level. 
(4) Real-world CVE. The dataset we collected contains real-world CVEs, which are more representative than synthetic vulnerabilities. 
(5) Large-scale. A vast volume of 30,987 vulnerability patches enables robust statistics and enhances machine learning via increased instances, far more than other benchmarks' limited samples. 

In summary, at the conceptual level, our dataset is more comprehensive and representative than existing ones.

\parh{Dataset coverage.}~As discussed in \S~\ref{subsec:dataset-analysis}, we further analyze the coverage of Common Weakness Enumeration (CWE) types in our dataset compared to other benchmarks. Our hypothesis is that a dataset exhibiting a wider range of CWE types will enable more comprehensive vulnerability modeling with greater potential for generalization across systems.

As shown in \T~\ref{tab:dataset}, our dataset covers more CWE types than all other benchmarks. 
Due to the limitation of space, we only show the total number of all CWE types across languages, but our dataset covers more CWE types even when focusing on one specific language.
For example, our dataset covers 134 CWE types in C/C++, while Juliet-C++ benchmark only covers 118 CWE types. Notably, it would be hard to estimate a clear number of potential CWEs in a specific language, since some of the CWEs are not language-specific. However, we can still observe that our dataset covers more CWE types than other benchmarks, which indicates that our dataset provides more comprehensive vulnerability coverage compared to existing benchmarks.

Beyond CWE coverage analysis, we also evaluate detected CWE coverage using a static analysis tool. The intuition is that a lower proportion of successfully detected vulnerabilities suggests a more challenging dataset, as existing flaws are harder to discover with standard tools. Such difficulty highlights the potential utility of advanced models. 
Specifically, we use Semgrep~\cite{semgrep}, a popular query-based static analysis tool, to detect potential vulnerabilities. We use its default ruleset, which contains 1,088 rules for Java, 655 rules for Python, and 133 rules for JS, to detect vulnerabilities in our dataset and other benchmarks. 
Defects4J only provides code patches, which makes analyzing it difficult. Consequently, we do not compare our benchmark with Defects4J.  
Juliet-C++ has a notably high detection rate at 35.7\%, presumably because it is a synthetic dataset.
LinuxFlaw and FUNDED, the benchmarks containing real-world CVEs,  have substantially lower detection rates of 2.2\% and 4.0\%, respectively.  
In contrast, our dataset has an even lower detection rate of 1.2\%, indicating that our dataset poses a more challenging problem. This is likely due to the fact that our dataset comprises more recent and intricate real-world vulnerabilities mined from up-to-date CVEs. Consequently, defects in our dataset may be more difficult to be detected using current approaches.

In summary, our collected dataset covers more CWE types and is more challenging for standard static analysis tools to assess, suggesting it is more comprehensive and representative than existing benchmarks. The broader range of hard-to-detect vulnerabilities in our dataset could support more robust vulnerability modeling and lead to repair systems with stronger generalization ability.

\parh{Dataset statistics.}~We first analyze the statistics of our collected dataset. As shown in \T~\ref{tab:ourdata}, our dataset contains 4466 vulnerabilities across seven languages. For each vulnerability, we report the average number of changed files, patch numbers, and changed lines of code, with results shown in the table. These results demonstrate that Java and C\# vulnerabilities are substantially more complex than those in the other languages, with average values for all metrics nearly double those of the other languages. This aligns with our expectation that these are usually served as Object-Oriented (OO) languages, and appear more compact in the coding specification.
Moreover, we observe that C language still consists of a large proportion of vulnerabilities, which indicates that (1) C language is still widely used in practice. (2) The unfamiliarity with the C language leads to ongoing vulnerability discovery. As an old language compared to Python, the language abstraction of C is relatively low, and developers need to manage memory manually (e.g. buffer overflows and memory leaks). This feature makes C language more error-prone, which leads to a large number of vulnerabilities.

\begin{table}[ht]
\caption{Statistics of our collected dataset.}
\resizebox{\linewidth}{!}{
    \begin{tabular}{l|l|l|l|l|l}
    \toprule
             Languages  & \# Case &\# Func & \# Avg diff file & \# Avg patch & \# Avg COL \\ \hline
               C++ &  411 & 2244 &2.88 & 5.46 & 86.81 \\
                C &  1575 & 6957 &2.14 & 4.42 & 62.97 \\
                Java &  541 & 6207 &5.74 & 11.47 & 297.13 \\
                Python &  863 & 5797 &3.26 & 6.72 & 113.2 \\
                JS &  636 & 5066 &4.26 & 7.97 & 130.32 \\
                Go &  355 & 3187 &4.54 & 8.98 & 195.43 \\
                C\# &  85 & 1529 &8.98 & 17.99 & 201.29 \\ \hline
                Total& 4466 & 30987 & 4.54 & 9.0 & 155.30 \\ 
                \bottomrule
    \end{tabular}
    }
    \label{tab:ourdata}    

\end{table}

\begin{figure}[!htbp]
    \includegraphics[width=0.9\linewidth]{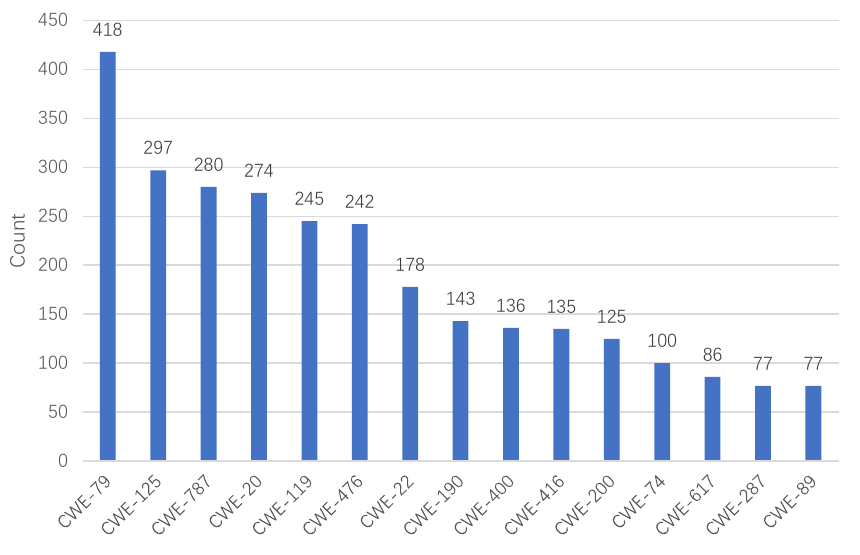}
    \caption{Top-15 CWE types of our dataset.}
    \label{fig:top15}
\end{figure}

Beyond analysis at the source code level, we also examine our dataset in terms of vulnerability types. As shown in \F~\ref{fig:top15}, the top 15 CWE types constitute over 55\% of our dataset, indicating it is comprehensive and representative. We also observe that Java and Python have similar CWE type distributions, while C/C++ has a distinct distribution.
The top five CWE types are CWE-79 (Cross-Site Scripting), CWE-125 (Out-of-bounds Read), CWE-787 (Out-of-bounds Write), CWE-20 (Improper Input Validation), and CWE-119 (Improper Restriction of Operations within the Bounds of a Memory Buffer). 
It is unsurprising that the dataset over-represents these CWE types, which are archetypical examples of memory-related vulnerabilities. Overall, these statistics suggest that our dataset effectively samples from the space of vulnerabilities and contains a diversity of complexity levels and types, especially for C/C++ and low-level languages.

\finding{1}{REEF, the framework we propose to collect the code vulnerabilities, considers more aspects than existing benchmarks, which in turn makes the dataset more practical to the real world. Through comprehensive analysis, we demonstrate that our dataset covers more CWE types and is more challenging than other benchmarks, indicating its comprehensiveness and representativity than other benchmarks.}

\subsection{RQ2: To what extent does the prompt design affect the generated message?}
\label{subsec:rq2}

We answer RQ2 from two aspects. First, we conduct a pilot study on C/C++ projects to analyze the generated message from the perspective of their basic understanding of vulnerabilities and fix records.
Second, we generate the message by using three commonly-used prompt patterns and compare them with the original message.

\parh{Pilot Study.}~Inspired by~\cite{ma2023oops}, we list the investigated LLMs in \T~\ref{tab:llm}. 
We use GPT-Neo~\cite{gao2020pile}, Llama-7B, and Llama-13B~\cite{touvron2023llama} fine-tuned on Alpaca~\cite{alpaca} and released by
LMFlow~\cite{lmflow}. Vicuna is Llama-based and fine-tuned on user-shared
conversations. ChatGLM~\cite{zeng2023glm} and Vicuna~\cite{vicuna2023} models use official code. We also include
commercial LLMs ChatGPT~\cite{chatgpt} and GPT-4~\cite{GPT4}. 
To make a fair comparison, we use the same prompt for all models. Moreover, we randomly select 15 CVE cases from C/C++ vulnerabilities.
After generating the message, we manually label the message as ``completely traceable'', ``somewhat traceable'' and ``non-traceable''.
Where if the message incorporates the information in the CVE patch or the vulnerability rationale, we label it as ``completely traceable''; if the message incorporates the information in the CVE patch or the vulnerability rationale partially, we label it as ``somewhat traceable''; otherwise, we label it as ``non-traceable''.

\begin{table}[t]
    \centering
    \caption{LLMs used in the experiment, with plausibility counts for ``completely
    traceable'', ``somewhat traceable'' and ``non-traceable'' labels. The symbol ``-" denotes that the corresponding statistic is unknown.} 
    \label{tab:llm}
    \resizebox{0.85\columnwidth}{!}{
    \begin{tabular}{l|c|c|c|c}
    \hline
    \textbf{Model} & \textbf{Vendor} & \textbf{Year} & \textbf{\# Para.} & \textbf{Traceability} \\ \hline
    GPT-Neo~\cite{gao2020pile} & EleutherAI & 2021 & 2.7B & 5/3/7 \\ \hline
    Llama-7B~\cite{touvron2023llama} & Meta & 2023 & 7B & 5/6/4 \\ \hline
    Llama-13B~\cite{touvron2023llama} & Meta & 2023 & 13B & 7/4/4 \\ \hline
    ChatGLM~\cite{zeng2023glm} & THU & 2023 & 6B & 10/4/1 \\ \hline
    Vicuna~\cite{vicuna2023} & BAIR & 2023 & 13B &  13/0/2 \\\hline
    ChatGPT~\cite{chatgpt} & OpenAI & 2022 & - & 13/1/1\\ \hline
    Tongyi~\cite{tyqw} & Alibaba & 2023  & - & 13/1/1\\ \hline
    GPT-4~\cite{GPT4} & OpenAI & 2023 & - & 15/0/0 \\ \hline
    \end{tabular}    
    }
    \end{table}

From the results, we observe that ChatGPT and Tongyi gain competitive performance in generating messages that are traceable to the CVE patch or the vulnerability rationale besides GPT-4.
Considering the limitations of APIs' accessibility and academic resources, we use Tongyi as the baseline model in the following experiments.

\parh{Prompt pattern comparison.}~Notably, we only generate one summarization message for each CVE, regardless of the number of commits and files. It could be the situation that one CVE contains multiple commits and files, which means the generated message is a summarization of all the commits and files.
However, since the input token length is limited, we set the maximum number of generated explanation tokens to 256. 
Once the input surpasses the limit, we truncate the input and generate the message based on the truncated input. Users can increase this number to generate more detailed explanations if needed.

\begin{figure}[t]
    \includegraphics[width=1.0\linewidth]{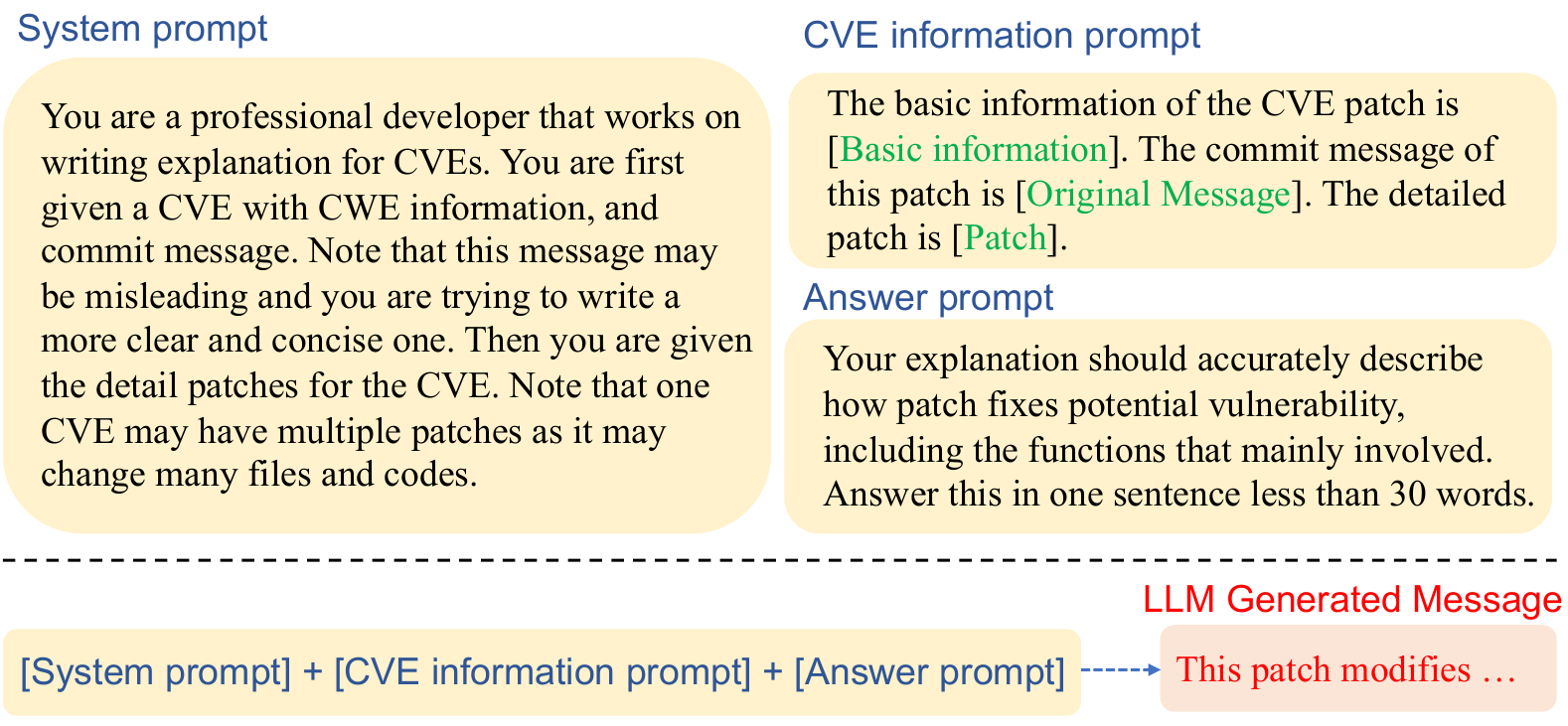}
    \caption{Example of the prompt we used to query  large language models.}
    \label{fig:example_prompt}
\end{figure}

Besides the token length, the prompt pattern is also a key point. It has been demonstrated that the output performance for code-related tasks is largely influenced by the type of prompt pattern~\cite{li2023feasibility}.
Specifically, we select 20 cases to generate messages based on the three prompt patterns, named ``Zero-shot'', ``One-shot'' and ``Few-shot''.

The prompts used to generate patch messages in the zero-shot setting are presented in Figure \ref{fig:example_prompt}. For one-shot and few-shot prompts, we give the large language model one or a few human-authored examples to guide the model in generating high-quality patch messages.
Here we do not consider the Chain-of-thought \cite{wei2022chain} pattern since no suitable rationale for code is available.
For each prompt pattern, we generate the message based on the CVE patch and the vulnerability rationale.
We then invite five experts to evaluate the generated message based on the following criteria:
\begin{itemize}
    \item \textbf{Comprehensiveness}: whether the message is comprehensive enough to explain the vulnerability and fix.
    \item \textbf{Consistency}: whether the message is consistent with the CVE patch and the vulnerability rationale.
    \item \textbf{Traceability}: whether the message is traceable to the CVE patch and the vulnerability rationale.
\end{itemize}

Each expert is asked to give a score range of 0 to 1 to demonstrate the generated quality regarding the corresponding criteria. After collecting the evaluation results, we calculate the average score for each prompt pattern. The results are shown in \T~\ref{tab:prompt-pattern}.
From the results, we observe that the ``Few-shot'' pattern achieves the highest score in all three criteria, while the ``One-shot'' pattern shares competitive results. As a trade-off between the 
performance and computation cost, we adopt ``One-shot'' as the default prompt pattern in this work.

\begin{table}[h]
    \centering
    \caption{Prompt pattern comparison.}
    \label{tab:prompt-pattern}
    \resizebox{0.85\columnwidth}{!}{
    \begin{tabular}{l|c|c|c}
    \hline
    \textbf{Prompt Pattern} & \textbf{Comprehensiveness} & \textbf{Consistency} & \textbf{Traceability} \\ \hline
    Zero-shot & 0.65 & 0.7 & 0.6 \\ \hline
    One-shot  & 1.0 & 0.95 & 0.95 \\ \hline
    Few-shot  & 1.0 & 1.0 & 0.95 \\ \hline
    \end{tabular}    
    }
\end{table}

\finding{2}{Large language models differ in their ability to generate vulnerability explanations traceable to source code and rationale. Tongyi and GPT models excelled in a pilot study, while a ``one-shot learning'' prompt achieved a satisfying performance for generating comprehensive, consistent, and traceable explanations in a prompt pattern comparison with an affordable query budget.}

\subsection{RQ3: How are generated messages in alignment with experts?}
\label{subsec:rq3}

To answer RQ3, we first present statistics comparing our generated messages to the original commit messages. We then conduct a human evaluation to assess the quality of our generated messages relative to the corresponding commit messages.

\parh{Statistics comparison}~We report statistics for our generated messages and the original commit messages in \T~\ref{tab:gener-msg}. On average, our generated messages contain 397.08 characters, 1.92 times more than the original commit messages. Moreover, the median length of our generated messages is 356, also higher than for the original commit messages. We attribute this to the fact that original commit messages are written by developers with varying perspectives and goals, yielding greater diversity than our generated messages.

Notably, some original commit messages are of low quality for two reasons:
(1) They are auto-filled by the GitHub platform. 
(2) They are too short (less than 20 characters.) 
We count the number of these low-quality commit messages and report them in the ``Lcmsg'' column of \T~\ref{tab:ourdata}.  Of the original commit messages in our dataset, nearly 5\% of messages are viewed as of low quality. They are unsuitable for providing informative vulnerability explanations and were thus excluded from our analyses, further motivating our work to generate comprehensive, high-quality vulnerability explanations.

The statistics demonstrate that our generated messages are substantially more detailed and consistent than the original commit messages. The higher word count suggests our messages provide more structured and in-depth vulnerability explanations overall compared to original commit messages, which are often quite brief and arbitrary. The filtering of low-quality messages is also prudent, as their inclusion could skew the statistics and make the dataset unsuitable for modeling. These results thus indicate we achieve our aim of generating high-quality, comprehensive vulnerability explanations.

\begin{table}[t]
\caption{Statistics of our generated message and the original commit message.}
    \begin{tabular}{l|l|l|l|l}
    \toprule
    Languages   & \# Case & \# Lcmsg  & \# Acmsg (Med.) & \# Agmsg (Med.) \\ \hline
    C++ &  411 & 21 & 234.93 (156) & 415.02 (364)  \\
    C &  1,575 & 122 & 380.0 (148) & 389.78 (351)  \\
    Java &  541 & 38 & 152.63 (68) & 399.51 (356)  \\
    Python &  863 & 36 & 204.11 (125) & 408.19 (363)  \\
    JS &  636 & 60 & 123.74 (57.0) & 382.84 (346.0)  \\
    Go &  355 & 20 & 237.68 (86) & 401.15 (376)  \\
    C\# &  85 & 3 & 109.85 (52) & 383.13 (340)  \\ \hline
    Total & 4,466 & 300 & 206.13 (98.85) & 397.08 (356.57) \\ 
    \bottomrule
\end{tabular}
    
    \label{tab:gener-msg}    

\end{table}

\parh{Human Study.}~We conduct a human evaluation to assess the quality of the generated messages. We recruit five experts, including two industrial developers and three academic researchers with expertise in software vulnerability detection, as participants. We randomly select 40 samples and create an online questionnaire for them. For each sample, we provide two messages without specifying their origin. On a scale of 1 to 5 (1 being completely unsatisfactory, 5 being fully satisfactory), participants score each message.
To ensure participants understand the task, we include five sanity check (SC) test items, considering only participants who answer all SC items correctly. Each participant evaluates 35 real samples and five SC items; we assign each sample to five participants. All participants pass the SC, taking an average of 35 minutes.

The human evaluation finds an average score of 3.05 for original messages and 3.70 for generated messages, a 21.31\% relative gain. Analysis of all responses shows that for 7.14\% of cases, the generated message seems worse, while for the remaining 92.86\%, the generated message is equal to or better than the original. The Fleiss' Kappa~\cite{fleiss1971measuring} of 0.92 indicates ``almost perfect agreement'' between participants.
These results suggest the generated messages are well-aligned with expert assessments and higher in quality than the original commit messages. The small minority of cases where the generated message seems inferior could be anomalous or reflects subtle aspects not captured in our message-generation approach. But the overwhelming expert preference for the generated messages, further evidenced by strong inter-rater agreement, indicates their quality is superior overall.

\finding{3}{Our generated vulnerability explanations are of superior quality to original commit messages according to both expert human evaluations and quantitative message statistics. Though a small fraction of generated messages are inferior, experts overwhelmingly prefer our generated explanations, indicating they are well-aligned with human assessments of explanation quality.}

\section{Discussion}
\label{sec:discussion}

\parh{Limitations and Threats to Validity.}~We now discuss the validity and limitations of this work. In this research, we collect the 
dataset from the real-world CVEs where the open-source projects are hosted on GitHub. However, not all open-source projects are hosted on GitHub, and we may miss some projects that are hosted on other platforms.
Moreover, since we only collect the CVEs from 2016, the previous CVEs are not included in our dataset. This may cause bias in the dataset, as some of the critical vulnerabilities may be discovered and fixed in previous years. 
In addition, we only collect the CVEs that are fixed by the developers. However, some CVEs are not fixed by the developers,
or the developers have already fixed the vulnerabilities but did not report or confirm the CVEs.
In summary, our collection framework tried to collect the CVEs as much as possible, but it is still possible that some CVEs are missing in our dataset, where future work can improve.

\parh{Message bias.}~In this work, we leverage the code understanding ability of large language models to generate additional messages for the collected vulnerabilities. Although we have specified strict rules as well as a dedicated designed prompt system, and our human evaluation shows that the generated messages are of high quality,
some of the generated messages may still be biased or unsatisfactory. For example, when the input code snippet is either too long or too short, the generated message may not be as good as the other cases. However, their superior ratings compared to original commit messages, as assessed by experts, indicate they still achieve the aim of producing informative vulnerability explanations, even if imperfect. Moreover, the quality of generated messages can be further improved with the help of more advanced language models.

\section{Conclusion}
\label{sec:conclusion}

In this paper, we propose a novel and practical approach to collect the real-world CVEs with detailed information automatically, 
and we leverage the code understanding ability of large language models to generate additional messages for the collected vulnerabilities.
Our framework successfully collects 4,466 CVEs from 2016 to 2023 and incorporates 30,987 messages for the collected patches. The collected dataset has been evaluated by the developers, and the results show that the generated messages are of high quality and can help the developers to understand the vulnerabilities. This work serves as a roadmap for researchers to construct better data-driven bug detection and auto-fix techniques.


\bibliographystyle{IEEEtran}
\bibliography{bib/IEEEabrv,bib/code,bib/sast,bib/cot,bib/literature,bib/zj}

\end{document}